\documentstyle[12pt,epsf]{article}

\begin{document}
\sloppy
\sloppy
\sloppy
$\ $
\begin{flushright}{UT-722,\ 1995}\end{flushright}
\vskip 1.5 truecm

\begin{center}
{\large{\bf BRST Symmetric Formulation of a Theory \\
 with Gribov-type  Copies }}
\end{center}
\vskip .75 truecm
\centerline{\bf Kazuo Fujikawa}
\vskip .4 truecm
\centerline {\it Department of Physics,University of Tokyo}
\centerline {\it Bunkyo-ku,Tokyo 113,Japan}
\vskip 1. truecm

\makeatletter
\@addtoreset{equation}{section}
\def\theequation{\thesection.\arabic{equation}}
\makeatother

\vskip 1. truecm

\begin{abstract}
A path integral with BRST symmetry can be formulated by summing the Gribov-type
 copies in a very specific way if the functional correspondence between $\tau$
and the gauge parameter $\omega$ defined by $\tau (x) = f( A_{\mu}^{\omega})$
is ``globally single valued'', where
$f( A_{\mu}^{\omega}) = 0 $ specifies the gauge condition.   A soluble gauge
model with Gribov-type copies recently analyzed  by Friedberg, Lee, Pang and
Ren satisfies this criterion.  A detailed BRST analysis of the soluble model
proposed by the above authors is presented. The BRST symmetry, if it is
consistently implemented, ensures the gauge independence of physical
quantities.
In particular, the vacuum (ground) state and the perturbative corrections to
the ground state energy in the above model are analysed from a view point of
BRST symmetry and $R_{\xi}$-gauge. Implications of the present analysis on some
aspects of the Gribov problem in non-Abelian gauge theory,  such as the $1/N$
expansion in QCD and also the dynamical instability of BRST symmetry, are
briefly discussed.

\par

\end{abstract}

\newpage
\section{Introduction}

\par
The quantization of non-Abelian gauge theory is complicated by the presence of
the so-called Gribov problem\cite{1}\cite{2}\cite{3}[4].
The Gribov problem in general suggests that the Coulomb gauge cannot completely
fix the gauge due to the presence of more than one gauge configurations which
satisfy the Coulomb gauge condition or , in
certain circumstances in compactified space-time, it even suggests the absence
of the gauge configuration which satisfies the Coulomb gauge condition [2]. In
Euclidean formulation of gauge theory, the Landau gauge also suffers from the
same complications. Although
the full details of the Gribov problem are not understood yet, a working
prescription is figured out if
one regards the Gribov problem as the appearance of several gauge copies( i.e.,
if one assumes that one can always find at least one gauge configuration which
satisfies the Coulomb gauge condition). In fact, it was noted some time
ago\cite{5} that if the functional correspondence between $\tau$ and the gauge
parameter $\omega$ defined by
\begin{equation}
\tau (x) =  \partial^{\mu}A_{\mu}^{\omega}(x)
\end{equation}
is ``globally single valued'', the path integral with BRST symmetry\cite{6} is
defined by summing over Gribov copies in a very specific way. For a more
general gauge fixing, (1.1) may be replaced by
\begin{equation}
\tau (x) =  f( A_{\mu}^{\omega}(x))
\end{equation}
where $f( A_{\mu}^{\omega}(x)) = 0$ specifies the gauge condition.
In (1.1) or (1.2)  $A_{\mu}^{\omega}(x)$ stands for the gauge field obtained
from $A_{\mu}(x)$ by a gauge transformation specified by the gauge orbit
parameter $\omega (x)$.
For an infinitesimal $\omega (x)$, one has
\begin{equation}
A_{\mu}^{a\omega}(x) = A_{\mu}^{a}(x) + \partial_{\mu}\omega^{a}(x)
                       - g f^{abc}A_{\mu}^{b}(x)\omega^{c}(x)
\end{equation}
One may regard (1.1) or (1.2) as a functional correspondence between $\tau$ and
$\omega$ parametrized by $A_{\mu}$.
The globally single valued correspondence between $\tau$ and $\omega$, which is
explained in more detail in Section 2, is required for any value of $A_{\mu}$
to write a simple path integral formula.
[ If one restricts oneself to only one of the Gribov copies by some
means\cite{7},
one can also incorporate the idea of BRST symmetry. But a simple
 prescription which selectively picks up only one copy  appears to be misssing
at this moment.]

 Recently a very detailed analysis of the Gribov problem was performed by
Friedberg, Lee ,Pang and Ren\cite{8} on the basis of a soluble gauge model
which exhibits Gribov-type copies. One of the main conclusions in \cite{8} is
that the singularity associated with the
so-called Gribov horizon is immaterial at least in their soluble
model and one may incorporate all the Gribov copies in a very specific way.
They showed how this prescription works in the
soluble model proposed by them.

The present work is motivated by the fact that the model and the gauge choice
in Ref.\cite{8} satisfy our criterion in (1.1) or (1.2). One can thus formulate
a BRST invariant path integral for the model proposed in [8] by summing over
Gribov-type copies ;\ the physical quantities thus calculated agree with those
in Ref.\cite{8}.
 Note that the BRST symmetry deals with an extended Hilbert space which
contains indefinite metric in general, although the
physical sector specified by BRST cohomology contains only positive metric.
Since the BRST symmetry plays a fundamental role in modern gauge theory, we
here present a detailed BRST analysis of the soluble model proposed in
\cite{8}.

\section{BRST invariant path integral in the presence of Gribov copies}
\par
In this Section we recapitulate the essence of the argument presented in
\cite{5}. We start with the Faddeev-Popov formulation of the Feynman-type gauge
condition \cite{9}\cite{10}. The vacuum-to-vacuum transition amplitude is
defined by
\begin{equation}
\langle +\infty|-\infty\rangle = {\int}{\cal D}A_{\mu}^{\omega}{\cal D}C
\delta(\partial^{\mu}A_{\mu}^{\omega} - C)\Delta(A)
exp\{iS(A_{\mu}^{\omega}) - \frac{i}{2\alpha}{\int}C(x)^{2}dx\}
\end{equation}
where $S(A_{\mu}^{\omega})$ stands for the action invariant under the
Yang-Mills local gauge transformation. The positive constant $\alpha$  is a
gauge fixing parameter which specifies the Feynman-type gauge condition. [The
equations
in this Section are written in the Minkowski metric, but they should really be
interpreted in the Euclidean metric to render the functional integral and the
Gribov problem well-defined.]
In the following we often suppress the internal symmetry indices, and instead
we write the gauge parameter explicitly: $A_{\mu}^{\omega}$ indicates the gauge
field which is obtained from $A_{\mu}$ by a gauge transformation specified by
$\omega(x)$. The determinant factor $\Delta(A)$ is defined by\cite{9}
\begin{eqnarray}
\Delta(A)^{-1}&=&{\int}{\cal D}\omega{\cal D}C\
\delta(\partial^{\mu}A_{\mu}^{\omega} - C)
exp\{ - \frac{i}{2\alpha}{\int}C(x)^{2}dx\}\nonumber\\
&\approx&const\{\sum_{k}|det[\frac{\partial}{\partial\omega_{k}}
\partial^{\mu}A_{\mu}^{{\omega}_{k}}]|^{-1}\}
\end{eqnarray}
where the summation runs over all the gauge equivalent configurations
satisfying $\partial^{\mu}A_{\mu}^{{\omega}_{k}}=0$ , which were
found by Gribov\cite{1} and others\cite{2}\cite{3}\cite{4}.
Equation(2.2) is valid only for sufficiently small $\alpha$, since
the parameter $\omega^{\prime}(\omega,A,C)$ defined by
\begin{equation}
\partial^{\mu}A_{\mu}^{\omega^{\prime}(\omega,A,C)}=
\partial^{\mu}A_{\mu}^{\omega}- C
\end{equation}
has a complicated branch structure for large $C$ in the presence of Gribov
ambiguities. Obviously the Feynman-type gauge formulation becomes even more
involved than the Landau-type gauge condition.

It was suggested in \cite{5} to replace equation (2.1) by
\begin{equation}
\langle +\infty|-\infty\rangle = \frac{1}{N}{\int}{\cal D}A_{\mu}^{\omega}{\cal
D}C
\delta(\partial^{\mu}A_{\mu}^{\omega} -
C)det[\frac{\partial}{\partial\omega}\partial^{\mu}A_{\mu}^{\omega}]
exp\{iS(A_{\mu}^{\omega}) - \frac{i}{2\alpha}{\int}C(x)^{2}dx\}
\end{equation}
The crucial difference between (2.1) and (2.4) is that (2.4) is local in the
gauge space $\omega(x)$ (i.e., the gauge fixing factor and the compensating
factor are defined at the identical $\omega$ ),whereas $\Delta(A)$ in (2.1) is
gauge independent and involves a non-local factor in $\omega$ as is shown in
(2.2). As the determinant in (2.4) depends on $A_{\mu}^{\omega}$, the entire
integrand in (2.4) is in general no more degenerate with respect to gauge
equivalent configurations even if the gauge fixing term itself may be
degenerate for certain configurations. Another important point is that one
takes the absolute values of determinant factors in (2.2) thanks to the
definition of the $\delta$-function, whereas just the determinant which can be
negative as well as positive appears in (2.4). It is easy to see that
(2.4) can be rewritten as
\begin{equation}
\langle +\infty|-\infty\rangle = \frac{1}{\tilde{N}}{\int}{\cal
D}A_{\mu}^{\omega}{\cal D}B{\cal D}\bar{c}{\cal D}c
\ exp\{iS(A_{\mu}^{\omega}) +i {\int}{\cal L}_{g}dx\}
\end{equation}
where
\begin{equation}
{\cal L}_{g}= -\partial^{\mu}B^{a}A_{\mu}^{a\omega} + i\partial^{\mu}
\bar{c}^{a}(\partial_{\mu}-gf^{abc}A_{\mu}^{b\omega})c^{c}
+ \frac{\alpha}{2}B^{a}B^{a}
\end{equation}
with $B^{a}$ the Lagrangian multiplier field, and $\bar{c}^{a}$ and
$c^{a}$ the (hermitian) Faddeev-Popov ghost fields; $f^{abc}$ is the
structure constant of the gauge group and $g$ is the gauge coupling constant.
If one imposes the hermiticity of $\bar{c}^{a}$ and $c^{a}$, the phase factor
of the determinant in (2.4) cannot be removed.
The normalization constant $\tilde{N}$  in (2.5)
includes the effect of Gaussian integral over $B$ in addition to $N$ in (2.4),
and in fact $\tilde{N}$ is independent of $\alpha$. See eq. (2.10).

This ${\cal L}_{g}$ as well as the starting gauge invariant Lagrangian are
invariant under the BRST transformation defined by
\begin{eqnarray}
\delta_{\theta} A_{\mu}^{a\omega} &=& i\theta [\partial_{\mu}c^{a} -gf^{abc}
A_{\mu}^{b\omega}c^{c}]\nonumber\\
\delta_{\theta} c^{a} &=& i\theta (g/2)f^{abc}c^{b}c^{c}\nonumber\\
\delta_{\theta} \bar{c}^{a} &=& \theta B^{a}\nonumber\\
\delta_{\theta} B^{a} &=& 0
\end{eqnarray}
where $\theta$  and  the ghost variables $c^{a}(x)$ and $\bar{c}^{a}(x)$ are
elements of the Grassmann algebra , i.e., $\theta^{2} = 0$. This transformation
can be confirmed to be nil-potent
$\delta^{2} = 0$, for example,
\begin{equation}
\delta_{\theta_{2}}(\delta_{\theta_{1}} A_{\mu}^{a\omega}) = 0
\end{equation}
One can also confirm that the path integral measure in (2.5) is invariant under
(2.7).
Note that the transformation (2.7) is ``local'' in the $\omega$ parameter;\
precisely for this property, the prescription in (2.4) was chosen.

To interprete  the path integral measure in (2.4) as the path integral  over
all the gauge field configurations divided by the gauge volume, namely
\begin{equation}
\langle +\infty|-\infty\rangle ={\int}\frac{{\cal D}A_{\mu}^{\omega}}{gauge\
volume(\omega)}
\ exp\{iS(A_{\mu}^{\omega})\}
\end{equation}
one needs to define the normalization factor in (2.4) by
\begin{eqnarray}
N &=& {\int}{\cal D}\omega{\cal D}C
\delta(\partial^{\mu}A_{\mu}^{\omega} -
C)det[\frac{\partial}{\partial\omega}\partial^{\mu}A_{\mu}^{\omega}]
exp\{- \frac{i}{2\alpha}{\int}C(x)^{2}dx\}\nonumber\\
  &=& {\int}{\cal D}\omega
det[\frac{\partial}{\partial\omega}\partial^{\mu}A_{\mu}^{\omega}]
exp\{-
\frac{i}{2\alpha}{\int}(\partial^{\mu}A_{\mu}^{\omega})^{2}dx\}\nonumber\\
  &=& {\int}{\cal D}\tau\ exp\{- \frac{i}{2\alpha}{\int}\tau (x)^{2}dx\}
\end{eqnarray}
where the function $\tau$ is defined by
\begin{equation}
\tau (x)\equiv \partial^{\mu}A_{\mu}^{\omega}(x)
\end{equation}
and the determinant factor is regarded as a Jacobian for the change
of variables from $\omega (x)$ to $\tau (x)$.
Although we use the Feynman-type gauge fixing (2.11) as a typical
example in this Section, one may replace (2.11) by
\begin{equation}
\tau (x)\equiv f( A_{\mu}^{\omega}(x))
\end{equation}
to deal with a more general gauge condition
\begin{equation}
 f( A_{\mu}^{\omega}(x)) = 0
\end{equation}
It is crucial to establish that the normalization factor in (2.10) is
independent of $A_{\mu}$. Only in this case, (2.4) defines an acceptable vacuum
transition amplitude. The Gribov ambiguity in the present
case appears as a non-unique correspondence between $\tau (x)$ and $\omega (x)$
in (2.11), as is schematically shown in Fig. 1 which includes 3 Gribov copies.
The path
integral in (2.10) is performed along the contour in Fig. 1. As the
Gaussian function is regular at any finite point, the complicated
contour in Fig. 1 gives rise to the same result in (2.10) as a
 contour corresponding to $A_{\mu} = 0$. In the present path integral
formulation, the evaluation of the normalization factor in (2.10) is the only
place where we explicitly encounter the multiple solutions of gauge fixing
condition.[ If the normalization factor $N$ should depend on gauge field
$A_{\mu}$, the factor $N$, which is gauge independent in the sense that we
integrated over entire gauge orbit, needs
to be taken inside the path integral in (2.4). In this case one looses the
simplicity of the formula (2.4).]

The basic assumption we have to make is therefore that (2.11) in the context of
the path integral (2.10) is ``globally single-valued'', in the sense that the
asymptotic functional correspondence between $\omega$ and $\tau$ is little
affected by a fixed $A_{\mu}$ with
 $\partial^{\mu}A_{\mu} = 0$\cite{5}; \ Fig.1 satisfies this requirement.
This assumption appears to be physically reasonable if the second derivative
term of the gauge orbit parameter dominates the functional
correspondence in (2.11),
though it has not been established mathematically. To define the functional
correspondence between $\omega$ and $\tau$ in (2.11), one needs in
general some notion of norm such as $L^{2}$-norm for which the Coulomb gauge
vacuum
is unique [3][4].  The functional configurations which are square integrable
however have zero measure in the path integral[11], and this makes the precise
analysis of (2.11) very complicated:\  At least what we need to do is to start
with an expansion of a generic
field variable in terms of some complete orthonormal basis set (which means
that the field is  inside the $L^{2}$-space)  and then let each expansion
coefficient vary from
$-\infty$ to $+\infty$(which means that the field is outside the
$L^{2}$-space).

If the Gribov problem simply means the situation as is schematically shown in
Fig. 1, the prescription in (2.4) may be justified. The indefinite signature of
the determinant factor in (2.4) is not a difficulty in the framework of
indefinite metric field theory\cite{12}\cite{13}  since  the determinant factor
is associated with the Faddeev-Popov ghost fields and the BRST cohomology
selects the positive definite physical space.  On the other hand , the Gribov
problem may aslo suggest  that one cannot bring the relation(2.11) with fixed
$A_{\mu}$ to $\partial^{\mu}A_{\mu}^{\omega}=0$ by any gauge
transformation\cite{2}. If this is the case, the asymptotic behavior of the
mapping (2.11) is in general modified by $A_{\mu}$ and our prescription cannot
be justified.  Consequently, the prescription (2.4) may be valid to the extent
that one can always achieve the condition $\partial^{\mu}A_{\mu} = 0$ by means
of suitable ( but not necessarily
unique) gauge transformations. We also note that the limit $\alpha \rightarrow
large$ in (2.5) and (2.10) corresponds to a very loose gauge fixing.

\section{BRST analysis of a soluble gauge model}
{\bf  3.1, A SOLUBLE GAUGE MODEL}
\par
The soluble gauge model of Friedberg, Lee, Pang and Ren\cite{8} is defined by
\begin{equation}
{\cal L}= \frac{1}{2}\{[\dot{X}(t)+g\xi(t)Y(t)]^{2} + [\dot{Y}(t) -
g\xi(t)X(t)]^{2} + [\dot{Z}(t) - \xi(t)]^{2}\} - U(X(t)^{2} +Y(t)^{2})
\end{equation}
where $\dot{X}(t)$, for example, means the time derivative of $X(t)$, and the
potential $U$ depends only on the combination $X^{2} + Y^{2}$. This Lagrangian
is invariant under a local gauge transformation
parametrized by $\omega (t)$ ,
\begin{eqnarray}
X^{\omega}(t) &=& X(t)\cos g\omega(t) - Y(t)\sin g\omega(t)\nonumber\\
Y^{\omega}(t) &=& X(t)\sin g\omega(t) + Y(t)\cos g\omega(t)\nonumber\\
Z^{\omega}(t) &=& Z(t) + \omega(t)\nonumber\\
\xi^{\omega}(t) &=& \xi(t) + \dot{\omega}(t)
\end{eqnarray}

The gauge condition ( an analogue of $A_{0} = 0$ gauge )
\begin{equation}
\xi(t) = 0
\end{equation}
or ( an analogue of $A_{3} = 0$ gauge )
\begin{equation}
Z(t) = 0
\end{equation}
is well-defined without suffering from Gribov-type copies. However,
it was shown in \cite{8} that the gauge condition ( an analogue of the Coulomb
gauge )
\begin{equation}
Z(t) - \lambda X(t) = 0
\end{equation}
with a constant $\lambda$ suffers from the Gribov- type complications. This is
seen by using the notation in (3.2) as
\begin{eqnarray}
Z^{\omega}(t) - \lambda X^{\omega}(t) &=& Z(t) + \omega(t) - \lambda
X(t)\cos g\omega(t) + \lambda Y(t)\sin g\omega(t)\nonumber\\
&=& \omega(t) +\lambda\sqrt{X^{2} +Y^{2}}[\cos \phi(t) -\cos (g\omega(t) +
\phi(t))] = 0
\end{eqnarray}
where we used the relation(3.5) and
\begin{equation}
X(t) = \sqrt{X^{2} +Y^{2}}\cos\phi(t), \ \ Y(t) =\sqrt{X^{2} +Y^{2}}\sin\phi(t)
\end{equation}
{}From a view point of gauge fixing, $\omega(t)= 0$ is a solution of
(3.6) if (3.5) is satisfied. By analyzing the crossing points of two
graphs in $(\omega, \eta) $ plane defined by
\begin{eqnarray}
\eta &=& \frac{1}{\lambda\sqrt{X^{2}+Y^{2}}}\omega\nonumber\\
\eta &=& \cos ( g\omega +\phi) -\cos\phi
\end{eqnarray}
one can confirm that eq.(3.6) in general has more than one solutions for
$\omega$.

{}From a view point of general gauge fixing procedure, we here regard the
algebraic gauge fixing such as (3.3) and (3.4) well-defined; \ in the analysis
of the Gribov problem in Ref.[2], the algebraic gauge fixing [3] is excluded.

The authors in Ref.\cite{8} started with the Hamiltonian formulated
in terms of  the well-defined gauge $\xi (t) = 0 $ in (3.3) and then
faithfully rewrote the Hamiltonian in terms of the variables defined by the
``Coulomb gauge'' in (3.5). By this way, the authors in \cite{8} analyzed in
detail the problem related to the Gribov copies and the so-called Gribov
horizons where the Faddeev-Popov determinant
vanishes. They thus arrived at a prescription which sums over all the
Gribov-type copies in a very specific way. As is clear from their derivation,
their specification satisfies the unitarity and gauge independence.

In the context of  BRST invariant path integral discussed in Section 2, the
crucial relation (2.11) becomes
\begin{eqnarray}
\tau (t) &=& Z^{\omega}(t) - \lambda X^{\omega}(t)\nonumber\\
         &=& \omega(t) + Z(t) -\lambda X(t)\cos g\omega(t)  +\lambda Y(t)\sin
g\omega(t)
\end{eqnarray}
in the present model. For $X = Y =0$ , the functional correspondence between
$\omega$ and $\tau$   is one-to-one and monotonous for any
fixed value of $t$ . When one varies $X(t), Y(t)$ and $Z(t)$ continuously, one
deforms this monotonous curve continuously. But the asymptotic correspondence
between $\omega (t)$ and $\tau (t)$ at $\omega(t) = \pm\infty$ for each value
of $t$ is still kept preserved, at
least for any fixed $X(t), Y(t)$ and $Z(t)$ . This correspondence
between $\omega (t)$ and  $\tau (t)$ thus satisfies our criterion
discussed in connection with (2.11). The absence of terms which contain the
derivatives of $\omega (t)$ in (3.9) makes the functional correspondence in
(3.9) well-defined and transparent.

{}From a view point of gauge fixing in (3.6), this ``globally single-valued''
correspondence between $\omega$ and $\tau$ means that one always obtains an
{\em odd} number of solutions for (3.6).
The prescription in \cite{8} is then viewed as a sum of all these solutions
with signature factors specified by the signature of the Faddeev-Popov
determinant
\begin{equation}
det\{\frac{\partial}{\partial\omega (t^{\prime})}[Z^{\omega}(t)-\lambda
X^{\omega}(t)]\} = det\{[1 + \lambda g Y^{\omega}(t)]
\delta(t-t^{\prime})\}
\end{equation}
evaluated at the point of solutions , $\omega = \omega(\sqrt{X^{2}+Y^{2}},
\phi)$, of (3.6). The row and column indices of the matrix in (3.10) are
specified by $t$ and $t^{\prime}$, respectively.
In the context of BRST invariant formulation, a pair-wise
cancellation of Gribov-type  copies takes place,
except for one solution, in the calculation of the normalization
factor in (2.10) or (3.21) below.

\noindent
{\bf 3.2, BRST INVARIANT PATH INTEGRAL}

The relation (3.9) satisfies our criterion discussed in connection with (2.11).
We can thus define an analogue of (2.5) for the Lagrangian (3.1) by
\begin{equation}
\langle +\infty|-\infty\rangle = \frac{1}{\tilde{N}}{\int}d\mu
\ exp\{iS(X^{\omega},Y^{\omega},Z^{\omega},\xi^{\omega}) +i {\int}{\cal
L}_{g}dt\}
\end{equation}
where
\begin{eqnarray}
S(X^{\omega},Y^{\omega},Z^{\omega},\xi^{\omega}) &=& {\int}{\cal
L}(X^{\omega},Y^{\omega},Z^{\omega},\xi^{\omega}) dt\nonumber\\
 &=& S(X,Y,Z,\xi)
\end{eqnarray}
in terms of the Lagrangian ${\cal L}$ in (3.1). The gauge fixing part of (3.11)
is defined by
\begin{equation}
{\cal L}_{g} = -\beta \dot{B} \xi^{\omega} + B(Z^{\omega} - \lambda X^{\omega})
+ \beta i \dot{\bar{c}}\dot{c} -i\bar{c}( 1 + g\lambda Y^{\omega})c
+\frac{\alpha}{2}B^{2}
\end{equation}
where $\alpha,\beta$ and $\lambda$ are numerical constants, and $\bar{c}$ and
$c$  are (hermitian) Faddeev-Popov ghost fields. $B$ is a Lagrangian multiplier
field. Note that ${\cal L}_{g}$ is hermitian. The integral measure in (3.11) is
given by
\begin{equation}
d \mu = {\cal D}X^{\omega}{\cal D}Y^{\omega}{\cal D}Z^{\omega}{\cal
D}\xi^{\omega}{\cal D}B{\cal D}\bar{c}{\cal D}c
\end{equation}
The Lagrangians ${\cal L}$ and ${\cal L}_{g}$ and the path integral measure
(3.14) are invariant under the BRST transformation defined by
\begin{eqnarray}
X^{\omega}(t,\theta)&=& X^{\omega}(t) -i\theta g c(t) Y^{\omega}(t)\nonumber\\
Y^{\omega}(t,\theta)&=& Y^{\omega}(t) +i\theta g c(t) X^{\omega}(t)\nonumber\\
Z^{\omega}(t,\theta)&=& Z^{\omega}(t) +i\theta c(t)\nonumber\\
\xi^{\omega}(t,\theta)&=& \xi^{\omega}(t) +i\theta \dot{c}(t)\nonumber\\
c(t,\theta) &=& c(t)\nonumber\\
\bar{c}(t,\theta) &=& \bar{c}(t) + \theta B(t)
\end{eqnarray}
where the parameter $\theta$ is a Grassmann number, $\theta^{2} = 0$.
Note that $\theta$ and ghost variables anti-commute.
In (3.15) we used a BRST superfield notation: \ In this notation, the second
component of a superfield proportional to $\theta$ stands for the BRST
transformed field of the first component. The second component is invariant
under BRST transformation which ensures the nil-potency of the BRST charge. In
the operator notation to be defined later, one can write , for example,
\begin{equation}
X^{\omega}(t,\theta) = e^{-\theta Q} X^{\omega}(t,0) e^{\theta Q}
\end{equation}
with a nil-potent BRST charge $Q$, $\{ Q,Q\}_{+} = 0$. Namely, the BRST
transformation is a translation in $\theta$-space, and $\theta Q$ is analogous
to momentum operator.

In (3.11)$\sim$ (3.15), we explicitly wrote the gauge parameter
$\omega$
to emphasize that BRST transformation is ``local'' in the $\omega$-space. For
${\cal L}_{g}$ in (3.13), the relation (3.9) is replaced by ( an analogue of
the Landau gauge )
\begin{eqnarray}
\tau (t) &\equiv& \beta \dot{\xi}^{\omega}(t) + Z^{\omega}(t) - \lambda
X^{\omega}(t)\nonumber\\
&=& \beta \ddot\omega (t) + \omega(t) + \beta\dot{\xi}(t)
+ Z(t) -\lambda X(t)\cos g\omega(t)  +\lambda Y(t)\sin g\omega(t)
\end{eqnarray}
 The functional correspondence between $\omega$ and $\tau$ is monotonous and
one-to-one for weak $\xi(t), Z(t), X(t) $ and $Y(t)$ fields;
\ this is understood if one rewrites the relation (3.17) for Euclidean time $t
= - it_{E}$ by neglecting weak fields as
\begin{displaymath}
\tau (t_{E}) = ( - \beta\frac{d^{2}}{dt_{E}^{2}} + 1 ) \omega(t_{E})
\end{displaymath}
The Fourier transform of this relation gives a one-to-one monotonous
correspondence between the Fourier coefficients of $\tau$ and $\omega$ for
non-negative $\beta$. The asymptotic functional correspondence between $\omega$
and $\tau$ for weak field cases is preserved even for any fixed strong fields
$\xi(t), Z(t), X(t) $ and $Y(t)$ for non-negative $\beta$ to the extent that
the term linear in $\omega (t)$
dominates the cosine and sine terms.  The correspondence between $\tau$ and
$\omega$ in (3.17) is quite complicated for finite $\omega (t)$ due to the
presence of the derivatives of $\omega(t)$.

Thus (3.17) satisfies our criterion of BRST invariant path integral for any
non-negative $\beta$.
The relation (3.9) is recovered if one sets $\beta = 0$ in (3.17);\ the
non-zero parameter $\beta \neq 0$ however renders a canonical structure of the
theory
better-defined. For example, the kinetic term for ghost fields in (3.13)
disappears for $\beta = 0$. In this respect the gauge (3.5) is
also analogous to the unitary gauge.
In the following we set $\beta = \alpha > 0$ in (3.13),
\begin{equation}
{\cal L}_{g} = -\alpha \dot{B} \xi^{\omega} + B(Z^{\omega} - \lambda
X^{\omega})
+ \alpha i \dot{\bar{c}}\dot{c} -i\bar{c}( 1 + g\lambda Y^{\omega})c
+\frac{\alpha}{2}B^{2}
\end{equation}
and let $\alpha \rightarrow 0$ later. In the limit $\alpha = 0$, one recovers
the gauge condition (3.5)  defined in Ref.\cite{8}.
This procedure is analogous to $R_{\xi}$-gauge ( or the $\xi$-limiting
process of Lee and Yang\cite{14} )\cite{15}, where the (singular) unitary
gauge is defined in the vanishing limit of the gauge parameter, $\xi
\rightarrow 0$:\ In (3.18) the parameter $\alpha$ plays the role of $\xi$ in
$R_{\xi}$-
gauge. In $R_{\xi}$-gauge one can keep $\alpha \neq 0$ without spoiling gauge
invariance, and the advantage of this approach with $\alpha \neq 0$ is that one
can avoid the appearance of (operator ordering) correction terms \cite{8}
when one moves from the Hamiltonian formalism to the Lagrangian formalism and
vice versa. This point will be discussed in detail when we analyze perturbative
corrections to the ground state energy in Section 4.

By using the BRST invariance, one can show the $\lambda$- independence
of (3.11) as follows:
\begin{equation}
\langle +\infty|-\infty\rangle_{\lambda+\delta \lambda} =
\langle +\infty|-\infty\rangle_{\lambda } -
\delta\lambda \frac{1}{\tilde{N}}{\int}d\mu [B(t)X^{\omega}(t) +
ig\bar{c}(t)Y^{\omega}(t)c(t)]
\ exp\{i {\int}({\cal L}+{\cal L}_{g})dt\}
\end{equation}
where we perturbatively expanded in the variation of ${\cal L}_{g}$  for a
change of the parameter $\lambda +\delta\lambda$ ,
\begin{equation}
{\cal L}_{g}(\lambda +\delta\lambda) = {\cal L}_{g}(\lambda)
- \delta\lambda [B(t)X^{\omega}(t) + ig\bar{c}(t)Y^{\omega}(t)c(t)]
\end{equation}
This expansion is justified since the normalization factor defined by (see
eq.(2.10))
\begin{eqnarray}
N &=& {\int}{\cal D}\tau\ exp\{- \frac{i}{2\alpha}{\int}\tau (x)^{2}dt\}
\end{eqnarray}
is independent of $\lambda$ provided that the global single-valuedness in
(3.17) is satisfied. [See also the discussion related to eq.(5.1).]  As was
noted before, this path integral for $N$, which depends on $\alpha$, is the
only place where we explicitly encounter the Gribov-type copies in the present
approach.
By denoting the BRST transformed variables by prime, for example,
\begin{equation}
X^{\omega}(t)^{\prime} = X^{\omega}(t) -
ig\theta c(t)Y^{\omega}(t)
\end{equation}
we have a BRST identity ( or Slavnov-Taylor identity [16])
\begin{eqnarray}
\lefteqn{\frac{1}{\tilde{N}}{\int}d\mu \bar{c}(t)X^{\omega}(t)
\ exp\{i {\int}({\cal L}+{\cal L}_{g})dt\}}\nonumber\\
&=&
\frac{1}{\tilde{N}}{\int}d\mu^{\prime}
\bar{c}(t)^{\prime}X^{\omega}(t)^{\prime}
\ exp\{i {\int}({\cal L}^{\prime}+{\cal L}_{g}^{\prime})dt\}\nonumber\\
&=&
\frac{1}{\tilde{N}}{\int}d\mu \bar{c}(t)X^{\omega}(t)
\ exp\{i {\int}({\cal L}+{\cal L}_{g})dt\}\nonumber\\
&+&
\theta\frac{1}{\tilde{N}}{\int}d\mu [B(t)X^{\omega}(t) +
ig\bar{c}(t)Y^{\omega}(t)c(t)]
\ exp\{i {\int}({\cal L}+{\cal L}_{g})dt\}
\end{eqnarray}
where the first equality holds since the path integral is independent of the
naming of integration variables provided that the asymptotic
behavior and the boundary conditions are not modified by the change
of variables.  The second equality in (3.23) holds due to the BRST
invariance of the measure and the action
\begin{eqnarray}
d\mu^{\prime} &=& d\mu,\nonumber\\
{\cal L}^{\prime} +{\cal L}_{g}^{\prime} &=& {\cal L} + {\cal L}_{g}
\end{eqnarray}
but
\begin{equation}
\bar{c}(t)^{\prime}X^{\omega}(t)^{\prime}
=\bar{c}(t)X^{\omega}(t)
+\theta [B(t)X^{\omega}(t) + ig\bar{c}(t)Y^{\omega}(t)c(t)]
\end{equation}
{}From (3.23) one concludes
\begin{equation}
\frac{1}{\tilde{N}}{\int}d\mu [B(t)X^{\omega}(t) +
ig\bar{c}(t)Y^{\omega}(t)c(t)]
\ exp\{i {\int}({\cal L}+{\cal L}_{g})dt\} =0
\end{equation}
and thus
\begin{equation}
\langle +\infty|-\infty\rangle_{\lambda+\delta \lambda} =
\langle +\infty|-\infty\rangle_{\lambda}
\end{equation}
in (3.19). This relation shows that the ground state energy is
independent of the parameter $\lambda$;\ in particular one can choose
$\lambda=0$ in evaluating the ground state energy, which leads to the gauge
condition (3.4) without Gribov complications.

In the path integral (3.11), one may impose {\em periodic}
boundary conditions in time $t$ on all the integration variables
and let the time interval $\rightarrow \infty$ later so
that BRST transformation (3.15) be consistent with the boundary
conditions.

The analysis in this sub-section is general but formal. In the next
sub-section we convert the path integral (3.11) to an operator Hamiltonian
formalism and analyze in detail the structure of the ground state and the gauge
independence of physical energy spectrum.

\noindent
{\bf 3.3, BRST ANALYSIS:\ OPERATOR HAMILTONIAN FORMALISM}\\
We start with the BRST invariant effective Lagrangian
\begin{eqnarray}
{\cal L}_{eff} &=& {\cal L} + {\cal L}_{g}\nonumber\\
         &=&
\frac{1}{2}\{[\dot{X}^{\omega}(t)+g\xi(t)^{\omega}Y^{\omega}(t)]^{2} +
[\dot{Y}^{\omega}(t) -
g\xi^{\omega}(t)X^{\omega}(t)]^{2} + [\dot{Z}^{\omega}(t) -
\xi^{\omega}(t)]^{2}\}\nonumber\\
&&- U[(X^{\omega}(t))^{2} +(Y^{\omega}(t))^{2}]\nonumber\\
&&-\alpha \dot{B}(t) \xi^{\omega}(t) + B(t)(Z^{\omega}(t) - \lambda
X^{\omega}(t))
+ \alpha i \dot{\bar{c}}(t)\dot{c}(t)\nonumber\\
&&-i\bar{c}(t)[ 1 + g\lambda Y^{\omega}(t)]c(t)
+\frac{\alpha}{2}B(t)^{2}
\end{eqnarray}
obtained from (3.1) and (3.18). A justification of (3.28), in particular its
treatment of Gribov-type copies, rests on the path integral representation
(3.11). In the following we suppress the suffix $\omega$, which emphasizes that
the BRST transformation is local in $\omega$-space.

One can construct a Hamiltonian from (3.28) as
\begin{eqnarray}
H&=& \frac{1}{2}[P_{X}^{2} +P_{Y}^{2} +P_{Z}^{2}] + U(X^{2} +Y^{2})\nonumber\\
  && + \xi G - B(Z-\lambda X) + i\frac{1}{\alpha}p_{c}p_{\bar{c}}
     + i\bar{c}( 1+\lambda gY)c - \frac{1}{2}\alpha B^{2}
\end{eqnarray}
where
\begin{eqnarray}
P_{X} &=& \frac{\partial}{\partial\dot{X}}{\cal L}_{eff} = \dot{X} +g\xi
Y\nonumber\\
P_{Y} &=& \frac{\partial}{\partial\dot{Y}}{\cal L}_{eff} = \dot{Y} -g \xi
X\nonumber\\
P_{Z} &=& \frac{\partial}{\partial\dot{Z}}{\cal L}_{eff} = \dot{Z} -
\xi \nonumber\\
P_{B} &=& \frac{\partial}{\partial\dot{B}}{\cal L}_{eff} = - \alpha
\xi\nonumber\\
P_{\bar{c}} &=& \frac{\partial}{\partial\dot{\bar{c}}}{\cal L}_{eff} =
i\alpha\dot{c}\nonumber\\
P_{c} &=& \frac{\partial}{\partial\dot{c}}{\cal L}_{eff} =
-i\alpha\dot{\bar{c}}
\end{eqnarray}
and the Gauss operator is given by
\begin{equation}
G \equiv g(XP_{Y} - YP_{X}) + P_{Z}
\end{equation}
We note that $(p_{\bar{c}})^{\dagger} = - p_{\bar{c}}$ and $(p_{c})^{\dagger} =
- p_{c}$ from (3.30). The quantization condition
\begin{equation}
[P_{B}, B] = \frac{1}{i}
\end{equation}
implies
\begin{equation}
[\xi, B] = \frac{i}{\alpha}
\end{equation}
and thus one may take the representation
\begin{equation}
B = \frac{1}{i\alpha}\frac{\partial}{\partial\xi}
\end{equation}
which is used later.

We also note the quantization conditions
\begin{eqnarray}
\{ p_{\bar{c}}, \bar{c}\}_{+} = \frac{1}{i} &\rightarrow&  \{\dot{c},
\bar{c}\}_{+} = -\frac{1}{\alpha},\nonumber\\
\{ p_{c}, c\}_{+} = \frac{1}{i} &\rightarrow&
 \{c, \dot{\bar{c}}\}_{+} = \frac{1}{\alpha}
\end{eqnarray}
The BRST charge is obtained from ${\cal L}_{eff}$ (3.28) via the
Noether current as
\begin{equation}
Q = cG - ip_{\bar{c}}B
\end{equation}
The BRST charge $Q$ is hermitian $Q^{\dagger} = Q$ and nil-potent
\begin{equation}
\{Q, Q\}_{+} = 0
\end{equation}
by noting $\{c, c\}_{+} = \{p_{\bar{c}}, p_{\bar{c}}\}_{+}= \{p_{\bar{c}},
c\}_{+} =0$. The BRST transformation (3.15) is generated  by $Q$, for example,
\begin{eqnarray}
e^{-\theta Q}X(t)e^{\theta Q} &=& X(t) - [\theta Q, X(t)]\nonumber\\
            &=& X(t) - i\theta g c(t)Y(t),\nonumber\\
e^{-\theta Q}\bar{c}(t)e^{\theta Q} &=& \bar{c}(t) - [\theta Q,
\bar{c}(t)]\nonumber\\
            &=& \bar{c}(t) + \theta B(t)
\end{eqnarray}
by noting $\theta^{2} = 0$.

Some of the BRST invariant \underline{physical} operators are given
by
\begin{eqnarray}
&& X^{2} + Y^{2},\ \   P_{X}^{2} +P_{Y}^{2},\nonumber\\
&& L_{Z} = XP_{Y} - YP_{X}
\end{eqnarray}
The Hamiltonian in (3.29) is rewritten by using the BRST charge as
\begin{equation}
H = H_{0} + i\{ Q, \xi p_{c}\}_{+} + \{ Q, \bar{c}(Z - \lambda X\}_{+} +
\frac{\alpha}{2}\{ Q, B\bar{c}\}_{+}
\end{equation}
with
\begin{equation}
H_{0}\equiv \frac{1}{2}[  P_{X}^{2} + P_{Y}^{2} + P_{Z}^{2}] +
            U( X^{2} + Y^{2} )
\end{equation}

One defines a physical state $\Psi$ as an element of BRST cohomology
\begin{equation}
\Psi \in \ Ker\ Q/ Im\ Q
\end{equation}
namely
\begin{equation}
Q\Psi = 0
\end{equation}
but $\Psi$ is {\em not} written in a form $\Psi = Q\Phi$ with a non-vanishing
$\Phi$.

The time development of $\Psi$ is dictated by Schroedinger equation
\begin{equation}
i\frac{\partial}{\partial t}\Psi (t) = H\Psi (t)
\end{equation}
and thus
\begin{eqnarray}
\Psi (\delta t) &=& e^{-iH\delta t}\Psi (0)\nonumber\\
&=& \Psi (0) - i\delta tH\Psi (0)\nonumber\\
&=&\Psi (0) -i\delta tH_{0}\Psi (0)\nonumber\\
&& -i\delta tQ\{ i\xi p_{c} + \bar{c}( Z- \lambda X) +
\frac{\alpha}{2}B\bar{c}\}\Psi (0)\nonumber\\
&\simeq& \Psi (0) -i\delta tH_{0}\Psi (0)
\end{eqnarray}
in the sense of BRST cohomology by noting (3.40) and $Q\Psi (0) = 0$. Note that
the Hamiltonian is BRST invariant
\begin{equation}
[ Q, H_{0}] = [ Q, H ] = 0
\end{equation}
If one solves the time independent Schroedinger equation
\begin{equation}
H_{0}\Psi (0) = E\Psi (0)
\end{equation}
with $Q\Psi (0) = 0$, one obtains
\begin{eqnarray}
\Psi (0)^{\dagger}e^{-iHt}\Psi (0) &=& e^{-iEt}\Psi (0)^{\dagger}\Psi
(0)\nonumber\\
&=& e^{-iEt}
\end{eqnarray}
by noting $\Psi (0)^{\dagger}Q = 0$. The eigen-value equation (3.47) is gauge
independent and thus $E$ is formally gauge independent, but the value of $E$ is
constrained by $Q\Psi(0) = 0$ and thus we need a more detailed analysis.

The basic task in the present BRST approach is to construct physical states
$\Psi$ satisfying (3.42). We construct such physical states, in particular the
ground state, as Fock states. For an explicit construction of physical states,
we use the harmonic potential considered in Ref.\cite{8}
\begin{equation}
U( X^{2} +Y^{2} ) = \frac{\omega^{2}}{2}( X^{2} +Y^{2} )
\end{equation}
and we write $H$ in (3.29) as
\begin{eqnarray}
H&=& \frac{1}{2}[P_{X}^{2} +P_{Y}^{2}] + \frac{\omega^{2}}{2}( X^{2} +Y^{2} )
+ \frac{1}{2}(gL_{Z})^{2}\nonumber\\
  && + \frac{1}{2}(\tilde{P_{Z}} + \tilde{\xi})^{2}+ \frac{1}{2\alpha}Z^{2} -
\frac{1}{2}\tilde{\xi}^{2} - \frac{1}{2\alpha}(\alpha B + Z)^{2}\nonumber\\
&& + \frac{i}{\alpha}p_{c}p_{\bar{c}} +i\bar{c}c -\lambda\{ Q, \bar{c}X\}_{+}
\end{eqnarray}
with\
\begin{eqnarray}
\tilde{P_{Z}} &\equiv& P_{Z} + gL_{Z} = G\nonumber\\
\tilde{\xi}  &\equiv& \xi -  gL_{Z}
\end{eqnarray}
Eq.(3.50) indicates that the freedom associated with $Z$ has positive norm but
the freedom $\tilde{\xi}$ has negative norm.
We thus define
\begin{eqnarray}
X &=& \frac{1}{\sqrt{2\omega}}( a_{X} + a_{X}^{\dagger} ),\nonumber\\P_{X} &=&
i\sqrt{\frac{\omega}{2}}( - a_{X} + a_{X}^{\dagger} ),\nonumber\\
Y &=& \frac{1}{\sqrt{2\omega}}( a_{Y} + a_{Y}^{\dagger} ),\nonumber\\
P_{Y} &=& i\sqrt{\frac{\omega}{2}}( - a_{Y} + a_{Y}^{\dagger} ),\nonumber\\
Z &=& \frac{1}{\sqrt{2\nu}}( d  + {d}^{\dagger} ),\nonumber\\
\tilde{P}_{Z} &=& P_{Z} + gL_{Z} = i\sqrt{\frac{\nu}{2}}[(b - d) -
(b - d)^{\dagger}],\nonumber\\
\tilde{\xi} &=& \xi - gL_{Z} = (-i)\sqrt{\frac{\nu}{2}}( b - b^{\dagger}
),\nonumber\\
\alpha B + Z &=& \frac{1}{\sqrt{2\nu}}( b + b^{\dagger} ),\nonumber\\c &=&
\frac{1}{\sqrt{2}\mu}( \hat{c} + \hat{c}^{\dagger} ),
\nonumber\\
p_{\bar{c}} &=& \frac{\mu}{\sqrt{2}}( \hat{c} - \hat{c}^{\dagger} ),\nonumber\\
\bar{c} &=& \frac{1}{\sqrt{2}\mu}( \hat{\bar{c}} + \hat{\bar{c}}^{\dagger}
),\nonumber\\
p_{c} &=& \frac{\mu}{\sqrt{2}}( -\hat{\bar{c}} + \hat{\bar{c}}^{\dagger} )
\end{eqnarray}
with
\begin{equation}
\nu\ \equiv \ \sqrt{\frac{1}{\alpha}}, \ \ \  \mu = \sqrt{\frac{1}{\nu}}
\end{equation}
In (3.52) the operator expansion of $X, Y, c$, and $\bar{c}$ is the standard
one, but the expansion of $\tilde{P_{Z}}, \tilde{\xi}$ and $\alpha B + Z$ is
somewhat unconventional.

The canonical commutators are satisfied by postulating
\begin{displaymath}
[ a_{X}, a_{X}^{\dagger} ] = 1,\nonumber
\end{displaymath}
\begin{displaymath}
[ a_{Y}, a_{Y}^{\dagger} ] = 1,\nonumber
\end{displaymath}
\begin{displaymath}
[ d, d^{\dagger} ] = 1,\nonumber
\end{displaymath}
\begin{eqnarray}
[ \bar{b}, \bar{b}^{\dagger} ] &=& -1,\nonumber\\
\{ \hat{c}, \hat{\bar{c}}^{\dagger}\}_{+} &=& -i,\nonumber\\
\{ \hat{c}^{\dagger}, \hat{\bar{c}}\}_{+} &=& i
\end{eqnarray}
and all other commutators are vanishing. Here we defined
\begin{equation}
b  \equiv  \bar{b} -i\frac{1}{\sqrt{2\nu}}gL_{Z}
\end{equation}
so that
\begin{displaymath}
P_{Z} =  i\sqrt{\frac{\nu}{2}}[(\bar{b} - d) -
(\bar{b} - d)^{\dagger}],
\end{displaymath}
\begin{displaymath}
\xi = (-i)\sqrt{\frac{\nu}{2}}(\bar{b} - \bar{b}^{\dagger} ),
\end{displaymath}
\begin{displaymath}
\alpha B + Z = \frac{1}{\sqrt{2\nu}}( \bar{b} + \bar{b}^{\dagger} )
\end{displaymath}
The operator  $b$ also satisfies
\begin{equation}
[ b, b^{\dagger}] = -1
\end{equation}
The variables $\bar{b}$ in (3.54) and $b$ in (3.56) carry negative norm.
The Hamiltonian (3.50) is then written as
\begin{eqnarray}
H &=& \omega (a_{X}^{\dagger}a_{X} + a_{Y}^{\dagger}a_{Y} + 1) +
      \frac{1}{2}(gL_{Z})^{2}\nonumber\\
  && +\nu[d^{\dagger}d - b^{\dagger}b + 1] +
i\nu[\hat{\bar{c}}^{\dagger}\hat{c} - \hat{c}^{\dagger}\hat{\bar{c}} +
i]\nonumber\\
  && +\lambda\{Q, \bar{c}X\}_{+}
\end{eqnarray}
with the BRST charge
\begin{eqnarray}
Q &=& cG - ip_{\bar{c}}B\nonumber\\
  &=& i\nu[\hat{c}^{\dagger}(b - d) - \hat{c}(b - d)^{\dagger}]\nonumber\\
  &=& i\nu[\hat{c}^{\dagger}(\bar{b} - d) - \hat{c}(\bar{b} - d)^{\dagger}] +
\sqrt{\frac{\nu}{2}}(\hat{c}^{\dagger} + \hat{c})gL_{Z}
\end{eqnarray}
and
\begin{equation}
L_{Z} = XP_{Y} - YP_{X} = i(a_{X}^{\dagger}a_{Y} - a_{Y}^{\dagger}a_{X})
\end{equation}
The BRST charge in (3.58) is nil-potent
\begin{equation}
Q^{2} = {\nu}^{2}\hat{c}^{\dagger}\hat{c}[b - d, b^{\dagger}
- d^{\dagger}] = 0
\end{equation}
by noting (3.54) and (3.56).

We thus define the (physical) ground state at $t = 0$ by
\begin{eqnarray}
b|0\rangle &=& d|0\rangle = 0,\nonumber\\
\hat{c}|0\rangle &=& \hat{\bar{c}}|0\rangle = 0,\nonumber\\
a_{X}|0\rangle &=& a_{Y}|0\rangle = 0
\end{eqnarray}
which ensures
\begin{equation}
Q|0\rangle = 0
\end{equation}
The zero-point energy of $Z$ and $\xi$ and the zero-point energy of
$c$ and $\bar{c}$ in (3.57) cancel each other  for the state $|0\rangle$.
When one defines a unitary transformation [8]
\begin{eqnarray}
a_{X} &=& \frac{1}{\sqrt{2}}(\tilde{a}_{X} -i\tilde{a}_{Y})\nonumber\\
a_{Y} &=& \frac{1}{\sqrt{2}}(-i\tilde{a}_{X} +\tilde{a}_{Y})
\end{eqnarray}
we can write the physical part of $H$ in (3.57) as
\begin{eqnarray}
H_{phys} &\equiv& \omega (a_{X}^{\dagger}a_{X} + a_{Y}^{\dagger}a_{Y} + 1)
       +\frac{1}{2}(gL_{Z})^{2}\nonumber\\
         &=& \omega (\tilde{a}_{X}^{\dagger}\tilde{a}_{X} + \tilde{a}
  _{Y}^{\dagger}\tilde{a}_{Y} + 1) +
\frac{g^{2}}{2}(\tilde{a}_{X}^{\dagger}\tilde{a}_{X} -
\tilde{a}_{Y}^{\dagger}\tilde{a}_{Y})^{2}
\end{eqnarray}

If one recalls the relation
\begin{eqnarray}
b &=& \frac{1}{\sqrt{2\nu}}[ \nu\frac{\partial}{i\partial \xi} +
      \nu Z + i\xi]\nonumber\\
b - d &=& \frac{1}{\sqrt{2\nu}}[ -i(P_{Z} + gL_{Z}) +
\nu\frac{\partial}{i\partial \xi}]
\end{eqnarray}
where we used  $\alpha B = \frac{\partial}{i\partial \xi}$ in (3.34), the state
$|0\rangle$ in (3.61) is required to satisfy
\begin{eqnarray}
\{ \nu\frac{\partial}{\partial \xi} -\xi + i \nu Z\}|0\rangle_{Z\xi} =
0,\nonumber\\
\{ \nu\frac{\partial}{\partial \xi} + \frac{1}{i}\frac{\partial}{\partial Z} +
gL_{Z}\}|0\rangle_{Z\xi} = 0
\end{eqnarray}
Eq.(3.66) has a solution
\begin{eqnarray}
|0,L_{Z}\rangle_{Z\xi} &=& N_{0}\exp\{\frac{1}{2\nu}\xi^{2} - i\xi Z
                                -igL_{Z}Z -\frac{\nu}{2}Z^{2}\}\nonumber\\
                &=& N_{0}\exp\{\frac{1}{2\nu}(\xi - i\nu Z)^{2} -igL_{Z}Z\}
\end{eqnarray}
which depends on $L_{Z}$\ ;\ $N_{0}$ is a normalization constant. The inner
product
of this state needs to be defined by means of a $90$ degree rotation in the
variable $\xi$
\begin{equation}
\langle 0,L_{Z}| 0,L_{Z}\rangle = N_{0}^{2}\int dZ d\xi e^{\frac{1}{\nu}\xi^{2}
                                                           - \nu Z^{2}}
\end{equation}
which reflects the fact that the $\xi$-variable carries  negative norm.
If one projects the state $|0,L_{Z}\rangle_{Z\xi}$ to the one with $p_{\xi} =
0$ by
Fourier transformation, which is the general procedure of Dirac in his
treatment of
singular Lagrangian [17], one obtains
\begin{equation}
\int d\xi |0,L_{Z}\rangle_{Z\xi} \sim e^{-igL_{Z}Z}
\end{equation}
This is the physical ground state naively expected for $(Z, P_{Z})$
system for a given $L_{Z}$  on the basis of invariance under the Gauss operator
in (3.31), and it is independent of the gauge parameter
$\nu = \sqrt{\frac{1}{\alpha}}$.

Other sectors of the ground state $|0\rangle_{XY\bar{c}c}$ are constructed in a
standard manner
\begin{equation}
\tilde{a}_{X}|0\rangle_{XY\bar{c}c} = \tilde{a}_{Y}|0\rangle_{XY\bar{c}c} =
\hat{c}|0\rangle_{XY\bar{c}c} = \hat{\bar{c}}|0\rangle_{XY\bar{c}c} = 0
\end{equation}
and the entire ground state in (3.62) is written as
\begin{equation}
|0\rangle = |0,L_{Z}\rangle_{Z\xi} \otimes |0\rangle_{XY\bar{c}c}
\end{equation}
where $L_{Z} = \tilde{a}_{X}^{\dagger}\tilde{a}_{X} -
\tilde{a}_{Y}^{\dagger}\tilde{a}_{Y}$ is replaced by a c-number by acting
$L_{Z}$ on $|0\rangle_{XY\bar{c}c}$.
In the present case, $L_{Z} =0 $.

The ground state thus defined has a time dependence described by  $H$ in (3.57)
\begin{eqnarray}
e^{-iH\delta t}|0\rangle &=& |0\rangle -i\omega\delta t|0\rangle + \delta
t\lambda Q\bar{c}X|0\rangle \nonumber\\
                         &\simeq& |0\rangle -i\omega\delta t|0\rangle
\end{eqnarray}
in the sense of BRST cohomology by noting (3.62). Thus we can {\em represent}
the vacuum (or ground)
state in $Ker\ Q/Im\ Q$ by $|0\rangle$ in  (3.71) for any time, and we have
\begin{equation}
\langle 0|e^{-iHt}|0\rangle = e^{-i\omega t}
\end{equation}

Excited physical states are represented by
\begin{equation}
\frac{1}{\sqrt{n_{1}!n_{2}!}}(\tilde{a}_{X}^{\dagger})^{n_{1}}
(\tilde{a}_{Y}^{\dagger})^{n_{2}}|0\rangle \equiv
|0,L_{Z}= n_{1}-n_{2}\rangle_{Z\xi} \otimes
\frac{1}{\sqrt{n_{1}!n_{2}!}}(\tilde{a}_{X}^{\dagger})^{n_{1}}(\tilde{a}_{Y}^{\dagger})^{n_{2}}|0\rangle_{XY\bar{c}c}
\end{equation}
One obtains the eigenvalue of $H$ in (3.57) and (3.64) as
\begin{eqnarray}
H\frac{1}{\sqrt{n_{1}!n_{2}!}}(\tilde{a}_{X}^{\dagger})^{n_{1}}
(\tilde{a}_{Y}^{\dagger})^{n_{2}}|0\rangle &=&
[\omega(n_{1} + n_{2} +1) + \frac{g^{2}}{2}(n_{1} - n_{2})^{2}]
\frac{1}{\sqrt{n_{1}!n_{2}!}}(\tilde{a}_{X}^{\dagger})^{n_{1}}
(\tilde{a}_{Y}^{\dagger})^{n_{2}}|0\rangle \nonumber\\
&& +\lambda Q\bar{c}X
\frac{1}{\sqrt{n_{1}!n_{2}!}}(\tilde{a}_{X}^{\dagger})^{n_{1}}
(\tilde{a}_{Y}^{\dagger})^{n_{2}}|0\rangle\nonumber\\
&\simeq&
[\omega(n_{1} + n_{2} +1) + \frac{g^{2}}{2}(n_{1} - n_{2})^{2}]
\frac{1}{\sqrt{n_{1}!n_{2}!}}(\tilde{a}_{X}^{\dagger})^{n_{1}}
(\tilde{a}_{Y}^{\dagger})^{n_{2}}|0\rangle \nonumber\\
\end{eqnarray}
in the sense of BRST cohomology, since
\begin{equation}
Q\frac{1}{\sqrt{n_{1}!n_{2}!}}(\tilde{a}_{X}^{\dagger})^{n_{1}}
(\tilde{a}_{Y}^{\dagger})^{n_{2}}|0\rangle = 0
\end{equation}
This (3.76) is confirmed by using the expression of $Q$ in (3.58) as
\begin{eqnarray}
&&e^{-\theta
Q}\frac{1}{\sqrt{n_{1}!n_{2}!}}(\tilde{a}_{X}^{\dagger})^{n_{1}}(\tilde{a}_{Y}^{\dagger})^{n_{2}}|0\rangle \nonumber\\
&&= e^{-\theta i\nu \hat{c}^{\dagger}
b}\frac{1}{\sqrt{n_{1}!n_{2}!}}(\tilde{a}_{X}^{\dagger})^{n_{1}}(\tilde{a}_{Y}^{\dagger})^{n_{2}}|0\rangle \nonumber\\
&&= e^{-\theta i\nu \hat{c}^{\dagger}
\bar{b}}|0,L_{Z}=n_{1}-n_{2}\rangle_{Z\xi} \otimes e^{-\theta
\sqrt{\frac{\nu}{2}}\hat{c}^{\dagger}
gL_{Z}}
\frac{1}{\sqrt{n_{1}!n_{2}!}}(\tilde{a}_{X}^{\dagger})^{n_{1}}(\tilde{a}_{Y}^{\dagger})^{n_{2}}|0\rangle_{XYc\bar{c}} \nonumber\\
&&= e^{\theta \sqrt{\frac{\nu}{2}} \hat{c}^{\dagger}g
(n_{1}-n_{2})}|0,L_{Z}=n_{1}-n_{2}\rangle_{Z\xi} \otimes e^{-\theta
\sqrt{\frac{\nu}{2}}\hat{c}^{\dagger}
g(n_{1}-n_{2})}
\frac{1}{\sqrt{n_{1}!n_{2}!}}(\tilde{a}_{X}^{\dagger})^{n_{1}}(\tilde{a}_{Y}^{\dagger})^{n_{2}}|0\rangle_{XYc\bar{c}} \nonumber\\
&&=
\frac{1}{\sqrt{n_{1}!n_{2}!}}(\tilde{a}_{X}^{\dagger})^{n_{1}}(\tilde{a}_{Y}^{\dagger})^{n_{2}}|0\rangle
\end{eqnarray}
where we used
\begin{eqnarray}
L_{Z}&=& \tilde{a}_{X}^{\dagger}\tilde{a}_{X} -
\tilde{a}_{Y}^{\dagger}\tilde{a}_{Y},\nonumber\\
b|0,L_{Z}=n_{1}-n_{2}\rangle_{Z\xi}&=& \{\bar{b}
-i\frac{1}{\sqrt{2\nu}}g(n_{1}-n_{2})\}|0,L_{Z}=n_{1}-n_{2}\rangle_{Z\xi}\nonumber\\
&=& 0
\end{eqnarray}
The second relation in (3.78) is regarded as a constraint on the ground state
of $\bar{b}$ for a given value of $L_{Z}$,which is a manifestation of the Gauss
constraint in the present formulation.

The states which include $b^{\dagger},d^{\dagger},\hat{c}^{\dagger}$,and
$\hat{\bar{c}}^{\dagger}$ excitations become unphysical and are removed by BRST
cohomology.  For example, the state
\begin{equation}
-\nu (b - d)^{\dagger}|0\rangle =  Q\hat{\bar{c}}^{\dagger}|0\rangle
\end{equation}
is BRST invariant and has energy (for $\lambda = 0$ )
\begin{equation}
HQ\hat{\bar{c}}^{\dagger}|0\rangle = (\omega +
\nu)Q\hat{\bar{c}}^{\dagger}|0\rangle
\end{equation}
but it is obviously excluded by BRST cohomology. The unphysical excitations
$b^{\dagger},d^{\dagger},\hat{c}^{\dagger}$,and $\hat{\bar{c}}^{\dagger}$ form
the components
of {\em non-trivial} BRST superfields $Z^{\omega}(t,\theta)$ and
$\bar{c}(t,\theta)$ in (3.15),
\begin{eqnarray}
Z^{\omega}(t,\theta) &=& Z^{\omega}(t) + i\theta c(t)\nonumber\\
                     &=& \frac{1}{\sqrt{2\nu}}( d + d^{\dagger})
                        + i\theta \frac{1}{\sqrt{2}\mu}(\hat{c} +
                          \hat{c}^{\dagger})\nonumber\\
\bar{c}(t,\theta) &=& \bar{c}(t) + \theta B(t)\nonumber\\
                  &=& \frac{1}{\sqrt{2}\mu}(\hat{\bar{c}} +
                       \hat{\bar{c}}^{\dagger}) + \theta\nu
                      \sqrt{\frac{\nu}{2}}[b - d + (b - d)^{\dagger}]
\end{eqnarray}
A characteristic property of these non-trivial superfields is that the second
components of the superfields, which are   BRST
transform of the first components, contain  terms $linear$ in the elementary
field. The basic  theorem of BRST symmetry is that any BRST
invariant state which contains those unphysical degrees of freedom
,$b^{\dagger}, d^{\dagger}, \hat{c}^{\dagger}$ and $\hat{\bar{c}}^{\dagger}$,
is written in the BRST exact form such as in (3.79) and thus
it is removed by BRST cohomology. See Ref.[13].

The present BRST analysis is in accord with an explicit construction of
physical states in Ref.\cite{8}. One can safely take the limit
$\alpha \rightarrow 0$ (or $\nu = \frac{1}{\sqrt{\alpha}} \rightarrow
\infty$) in the physical sector, though unphysical excitations such as in
(3.80) acquire infinite excitation energy in this limit just like unphysical
excitations in gauge theory defined by $R_{\xi}$-gauge\cite{15}.

\section{Perturbative calculation in path integral}

It has been shown in \cite{8} that the correction terms arising from operator
ordering plays a crucial role in the evaluation of perturbative corrections to
ground state energy in Lagrangian path integral
formula. This problem is often treated casually in conventional perturbative
calculations;\ a general belief ( and  hope) is that Lorentz invariance and
BRST invariance somehow takes care of the operator ordering problem. In the
following, we show that BRST invariance and $T^{\star}$-product prescription
reproduce the correct result of Ref.\cite{8} provided that one uses a
canonically well-defined gauge such as $R_{\xi}$-gauge with $\alpha \neq 0$ in
(3.18). This check is
important to establish the equivalence of (3.11) to the path integral formula
in Ref.\cite{8}. See also Ref.\cite{18}. If one starts with $\alpha = 0$ from
the on-set, one needs correction terms calculated in Ref.\cite{8}.

To be precise, what we want to evaluate is eq.(3.11), namely
\begin{equation}
\langle +\infty|-\infty\rangle = \frac{1}{\tilde{N}}{\int}d\mu
\ exp\{iS(X^{\omega},Y{\omega},Z^{\omega},\xi^{\omega}) +i {\int}{\cal
L}_{g}dt\}
\end{equation}
We define the path integral for a sufficiently large time interval
\begin{equation}
t\in [ T/2, -T/2 ]
\end{equation}
and let $T \rightarrow \infty$ later. In the actual calculation, there appear
two important aspects which need to be taken into account:\\
(i)\ We impose periodic boundary conditions on all the variables so that BRST
transformation (3.15) is well-defined including the boundary conditions.\\
(ii)\ In the actual evaluation of the path integral as well as Feynman diagrams
, we may apply the Wick rotation and perform Euclidean calculations.\\

The exact ground state energy of (4.1) is given by eq.(3.73) as
\begin{equation}
E = \omega
\end{equation}
Namely, we have no correction depending on the gauge parameter $\lambda$ and
the coupling constant $g$. As was already shown in (3.27), the absence of
$\lambda$ dependence is a result of BRST symmetry. This property is thus more
general and, in fact, it holds for all the energy spectrum of physical states;\
this can be shown by using the Schwinger's action principle [18] and the
definition of physical states in (3.42). The perturbative check of
$\lambda$-independence or Slavnov-Taylor identities in general is carried out
in the standard manner.
On the other hand, the absence of $g$- dependence is an effect of more
dynamical origin. In this Section we concentrate on the evaluation of
$g$-dependence by taking a view that the $\lambda$- independence has been
generally established in (3.27).

We first perform the Gaussian path integral over $B$-variable by noting
\begin{eqnarray}
{\cal L}_{gauge} &=& \frac{\alpha}{2}B^{2} + \alpha B\dot{\xi} + B(Z - \lambda
X)\nonumber\\
&=& \frac{\alpha}{2}[B + \frac{1}{\alpha}(\alpha\dot{\xi} + Z - \lambda X)]^{2}
- \frac{1}{2\alpha}[\alpha\dot{\xi} + Z - \lambda X]^{2}\nonumber\\
&\Rightarrow& - \frac{1}{2\alpha}[\alpha\dot{\xi} + Z - \lambda
X]^{2}\end{eqnarray}
We thus consider the path integral
\begin{equation}
\langle +\infty|-\infty\rangle = \frac{1}{N}{\int}{\cal D}X{\cal D}Y{\cal
D}Z{\cal D}\xi{\cal D}\bar{c}{\cal D}c
\ exp\{i {\int}{\cal L}_{eff}dt\}
\end{equation}
where $N$ is the original normalization constant in (3.21), and
\begin{eqnarray}
{\cal L}_{eff}
         &=& \frac{1}{2}\{[\dot{X}(t)+g\xi(t)Y(t)]^{2} + [\dot{Y}(t) -
g\xi(t)X(t)]^{2}\} - \frac{\omega^{2}}{2}[{X(t)}^{2} +{Y(t)}^{2}]
\nonumber\\
&&
+ \frac{1}{2}{\dot{Z}(t)}^{2} - \frac{1}{2\alpha}[Z(t) - \lambda X(t)]^{2}
- \frac{\alpha}{2}{\dot{\xi}(t)}^{2} + \frac{1}{2}{\xi (t)}^{2}\nonumber\\
&&
+ \lambda\dot{\xi}(t)X(t) + \alpha i \dot{\bar{c}}(t)\dot{c}(t) -i\bar{c}(t)( 1
+ g\lambda Y(t))c(t) \nonumber\\
&\equiv&
{\cal L}_{0} + {\cal L}_{I}
\end{eqnarray}
with
\begin{eqnarray}
{\cal L}_{0} &\equiv&
         \frac{1}{2}[{\dot{X}(t)}^{2} + {\dot{Y}(t)}^{2}] -
\frac{\omega^{2}}{2}[{X^(t)}^{2} +{Y(t)}^{2}]
\nonumber\\
&&
+ \frac{1}{2}{\dot{Z}(t)}^{2} - \frac{1}{2\alpha}Z(t)^{2}
- \frac{\alpha}{2}{\dot{\xi}(t)}^{2} + \frac{1}{2}{\xi (t)}^{2}\nonumber\\
&&
+ \alpha i \dot{\bar{c}}(t)\dot{c}(t) -i\bar{c}(t)c(t),\\
{\cal L}_{I} &\equiv&
g\xi (t)[\dot{X}(t)Y(t) - \dot{Y}(t)X(t)] + \frac{1}{2}g^{2}{\xi
(t)}^{2}[X(t)^{2} + Y(t)^{2}]
\end{eqnarray}
where we set $\lambda = 0$ in the final expressions in (4.7) and (4.8).

The propagators are defined by ${\cal L}_{0}$ in (4.7) in the standard manner
as
\begin{eqnarray}
\langle T^{\star} X(t_{1})X(t_{2})\rangle &=& \langle T^{\star}
Y(t_{1})Y(t_{2})\rangle\nonumber\\
&=& \int \frac{dk}{2\pi} e^{ik(t_{1} - t_{2})} \frac{i}{k^{2} - \omega^{2} +
i\epsilon}\nonumber\\
\langle T^{\star} Z(t_{1})Z(t_{2})\rangle
&=& \int \frac{dk}{2\pi} e^{ik(t_{1} - t_{2})} \frac{i}{k^{2} - 1/\alpha +
i\epsilon}\nonumber\\
\langle T^{\star} \xi (t_{1})\xi (t_{2})\rangle
&=& \int \frac{dk}{2\pi} e^{ik(t_{1} - t_{2})} \frac{-i/\alpha}{k^{2} -
1/\alpha + i\epsilon}\nonumber\\
\langle T^{\star} \bar{c}(t_{1})c(t_{2})\rangle
&=& \int \frac{dk}{2\pi} e^{ik(t_{1} - t_{2})} \frac{- 1/\alpha}{k^{2} -
1/\alpha + i\epsilon}
\end{eqnarray}
where we took $T\rightarrow \infty$ limit in the evaluation of those
propagators; \ this procedure is justified for the evaluation of corrections to
the ground state energy since all the momentum integrations in (4.11) below are
well-convergent.

Up to the second order of perturbation in ${\cal L}_{I} $,we have
\begin{eqnarray}
\langle +\infty|-\infty\rangle &=& \frac{1}{N}{\int}d\mu
\ e^{i{\int}{\cal L}_{0}dt}\{ 1 + i\int dt \langle T^{\star}\frac{1}{2}g^{2}\xi
(t)^{2}[X(t)^{2} + Y(t)^{2}]\rangle \nonumber\\
&+& \frac{(i)^{2}}{2!}\int dt_{1}dt_{2}\langle T^{\star}g^{2}\xi (t_{1})\xi
(t_{2})[\dot{X}(t_{1})Y(t_{1}) - \dot{Y}(t_{1})X(t_{1})]\nonumber\\
&& \ \ \ \ \ \ \ \ \ \ \ \ \ \ \ \ \ \ \ \ \ \ \ \ \ \ \  \times
[\dot{X}(t_{2})Y(t_{2}) - \dot{Y}(t_{2})X(t_{2})]\rangle \}\nonumber\\
                          &=& \frac{1}{N}{\int}d\mu
\ e^{i{\int}{\cal L}_{0}dt}\{ 1 + ig^{2}\int dt \langle T^{\star}\xi (t)\xi
(t)\rangle \langle T^{\star}X(t)X(t)\rangle \nonumber\\
&+& (i)^{2}g^{2}\int dt_{1}dt_{2}\langle T^{\star}\xi (t_{1})\xi (t_{2})\rangle
\langle T^{\star}\dot{X}(t_{1})\dot{X}(t_{2})\rangle \langle
T^{\star}Y(t_{1})Y(t_{2})\rangle \nonumber\\
&-& (i)^{2}g^{2}\int dt_{1}dt_{2}\langle T^{\star}\xi (t_{1})\xi (t_{2})\rangle
\langle T^{\star}\dot{X}(t_{1})X(t_{2})\rangle \langle
T^{\star}Y(t_{1})\dot{Y}(t_{2})\rangle \}
\end{eqnarray}
by taking the symmetry in $X$ and $Y$ into account.

The terms of order $g^{2}$ in (4.10) are written by using the propagators in
(4.9) as
\begin{eqnarray}
&&(-\frac{i}{\alpha}g^{2})\int dt\int \frac{dk}{2\pi}\frac{1}{k^{2} +
1/\alpha}\int \frac{dl}{2\pi}\frac{1}{l^{2} + \omega^{2}}\nonumber\\
&&+(\frac{i}{\alpha}g^{2})\int dt\int \frac{dk}{2\pi}\frac{1}{k^{2} +
1/\alpha}\int \frac{dl}{2\pi}\frac{l^{2}}{l^{2} + \omega^{2}}\frac{1}{(k
+l)^{2} + \omega^{2}}\nonumber\\
&&+(\frac{i}{\alpha}g^{2})\int dt\int \frac{dk}{2\pi}\frac{1}{k^{2} +
1/\alpha}\int \frac{dl}{2\pi}\frac{1}{l^{2} + \omega^{2}}\frac{l(l +k)}{(k
+l)^{2} + \omega^{2}}\nonumber\\
&=&(-\frac{i}{\alpha}g^{2})T\int \frac{dk}{2\pi}\frac{1}{k^{2} + 1/\alpha}\int
\frac{dl}{2\pi}\frac{1}{l^{2} + \omega^{2}}\nonumber\\
&&+(\frac{i}{\alpha}g^{2})T\int \frac{dk}{2\pi}\frac{1}{k^{2} + 1/\alpha}\int
\frac{dl}{2\pi}\frac{1}{l^{2} + \omega^{2}}\nonumber\\
&&+(-\frac{i}{\alpha}g^{2})T\int \frac{dk}{2\pi}\frac{1}{k^{2} + 1/\alpha}\int
\frac{dl}{2\pi}\frac{\omega^{2}}{l^{2} + \omega^{2}}\frac{1}{(k +l)^{2} +
\omega^{2}}\nonumber\\
&&+(\frac{i}{\alpha}g^{2})T\int \frac{dk}{2\pi}\frac{1}{k^{2} + 1/\alpha}\int
\frac{dl}{2\pi}\frac{1}{l^{2} + \omega^{2}}\frac{l(l +k)}{(k +l)^{2} +
\omega^{2}}
\end{eqnarray}
after the Wick rotation. We here note that the $T^{\star}$-product prescription
is crucial in obtaining (4.11) ;\ the $T^{\star}$-product  commutes with the
time derivative operation, which is intuitively understood from the fact that
the basic path integration variables are field variables ( and not their time
derivatives) in the Lagrangian path integral formula [19]. In this approach,
the conventional  $T$-product is defined from $T^{\star}$-product via the
Bjorken-Johnson-Low prescription [20].

The first two terms in (4.11) cancel each other. The last two terms in (4.11)
can be evaluated as follows: \ In the third term in (4.11), one can rewrite the
integrand as
\begin{eqnarray}
\frac{\omega^{2}}{l^{2} + \omega^{2}}\frac{1}{(k +l)^{2} + \omega^{2}} &=&
(\frac{1}{2i})^{2}(\frac{-1}{l + i\omega} + \frac{1}{l - i\omega})(\frac{-1}{l
+ k + i\omega} + \frac{1}{l + k -i\omega})\nonumber\\
&\rightarrow& (\frac{1}{2i})^{2}\{\frac{-1}{l + i\omega}\frac{1}{l + k
-i\omega} + \frac{1}{l - i\omega}\frac{-1}{l + k + i\omega}\}\nonumber\\
&=& (\frac{1}{2})^{2}\{\frac{1}{l + i\omega}\frac{1}{l + k -i\omega} +
\frac{1}{l - i\omega}\frac{1}{l + k + i\omega}\}
\end{eqnarray}
since the poles displaced in the same side of the real axis do not contribute
to the integral over $\int dl$;\ the contour can be shrunk to zero without
encircling poles. Similarly, in the last term in (4.11),
\begin{eqnarray}
\frac{1}{l^{2} + \omega^{2}}\frac{l(l +k)}{(k +l)^{2} + \omega^{2}}
&=& (\frac{1}{2})^{2}(\frac{1}{l + i\omega} + \frac{1}{l - i\omega})(\frac{1}{l
+ k + i\omega} + \frac{1}{l + k -i\omega})\nonumber\\
&\rightarrow&
(\frac{1}{2})^{2}\{\frac{1}{l + i\omega}\frac{1}{l + k -i\omega} + \frac{1}{l -
i\omega}\frac{1}{l + k + i\omega}\}
\end{eqnarray}
The last two terms in (4.11) thus  cancel each other, and we obtain only the
lowest order contribution in (4.11)
\begin{eqnarray}
\langle +\infty|-\infty\rangle &=& \frac{1}{N}{\int}d\mu
\ e^{i{\int}{\cal L}_{0}dt}\nonumber\\
&=& const \times \frac{1}{[2i\sin (\omega T/2)]^{2}}\frac{1}{N}\frac{det
[\alpha\partial_{t}^{2} + 1]}{\sqrt{det [\partial_{t}^{2} + 1/\alpha] det
[\alpha\partial_{t}^{2} + 1]}}\nonumber\\
&=& const \times \frac{1}{[2i\sin (\omega T/2)]^{2}} \ \ \ for\ T\rightarrow
\infty,
\end{eqnarray}
where the determinant factors coming from $Z, \xi, \bar{c}$ and $c$ integration
 combined with the normalization constant $N$  cancel completely among
themselves. The last expression in (4.14) is a standard path integral of
harmonic oscillators $X$ and $Y$ with periodic boundary conditions [21], and it
may be expanded as
\begin{equation}
const \times \frac{1}{[2i\sin (\omega T/2)]^{2}} =  const \times e^{-i\omega
T}(1 + \sum_{n_{1}, n_{2} = 0}^{\infty} e^{-i\omega T(n_{1} + n_{2})})
\end{equation}
for $T\rightarrow \infty$, where the summation over the non-negative integers
$n_{1}$ and $n_{2}$ excludes the case $n_{1} = n_{2} = 0$.
To be precise, the $T$-dependence of the normalization constant $N$ needs to be
taken into account to obtain the last expression of (4.14) [21].
The ground state energy is then obtained from
\begin{eqnarray}
\langle +\infty|-\infty\rangle &=& \lim_{T\rightarrow \infty} \langle
0|e^{-iH[T/2 - (-T/2)]}|0\rangle \nonumber\\
&=& \lim_{T\rightarrow \infty} const \times e^{-i\omega T}
\end{eqnarray}
which is justified for $T = -iT_{E}$ and $T_{E} \rightarrow \infty$ in
Euclidean theory. We thus obtain the ground state energy
\begin{equation}
E = \omega
\end{equation}
to be consistent with (3.73).

\section{Discussion and conclusion}

The BRST symmetry plays a central role in modern gauge theory,
and the BRST invariant path integral can be formulated by summing over all the
Gribov-type copies in a very specific manner provided that
the crucial correspondence in (2.11) or (3.17) is globally single valued[5].
This criterion is satisfied by the soluble gauge model proposed in Ref.[8], and
it is encouraging that the BRST invariant prescription is in accord with the
canonical analysis of the soluble gauge model in Ref.\cite{8}.
The detailed explicit analysis in Ref.\cite{8} and the present somewhat formal
BRST analysis are complementary to each other.
In Ref.\cite{8}, the problem related to the so-called Gribov horizon , in
particular the possible singularity associated with it, has been analyzed in
greater detail;\ this is crucial for the analysis of more general situation. On
the other hand, an advantage of the BRST analysis is that one can clearly see
the gauge independence of physical quantities such as the energy spectrum as a
result of BRST identity.

The BRST approach allows a transparent treatment of general class of gauge
conditions implemented by (3.18). This gauge condition with $\alpha \neq 0$
renders the canonical structure better-defined, and it allows simpler
perturbative treatments of the problems such as the corrections to the ground
state energy. Our calculation vis-a-vis the explicit canonical analysis in
Ref.\cite{8} may provide a (partial) justification of conventional covariant
perturbation theory in gauge theory, which is based on Lorentz invariance( or
$T^{\star}$-product ) and BRST invariance without the operator ordering terms.

Motivated by the observation in Ref.[8] to the effect  that the Gribov horizons
are not really singular in quantum mechanical sense, which is in accord
with our path integral in (3.21), we would like to make a
speculative comment on the role of Gribov copies in QCD.  First of
all, the topological phenomena related to instantons for which the
Coulomb gauge is generally singular  may be analyzed in the temporal gauge
$A_{0} = 0$;\ this gauge is relatively free from the Gribov complications[3]. A
semi-classical treatment of instantons
with small quantum fluctuations around them  will presumably give
 qualitatively reliable estimates of topological effects.

As for other non-perturbative effects such as quark confinement and
hadronic spectrum, one may follow the argument of Witten [22] on the basis of
$1/N$ expansion in QCD[23];\ he argues that the $1/N$ expansion scheme comes
closer to the real QCD than the instanton analysis in the study of hadron
spectrum. If this is the case, one can analyze  the qualitative aspects of
hadron spectrum on the basis of a sum of (an infinite number of) Feynman
diagrams. This diagramatic approach
or  an analytical treatment equivalent to it in the Feynman-type gauge deals
with topologically trivial gauge fields but may still suffer from the Gribov
copies, as is suggested by the analysis in Ref.[4]:
If one assumes that the vacuum is unique in this case as is the case in
$L^{2}$-space, the global single-valuedness in (2.11) in the context of path
integral (2.10) will be preserved for infinitesimally small fields $A_{\mu}$.
By a continuity argument, a smooth deformation of $A_{\mu}$ in $L^{2}$-space (
or its extension as explained in Section 2) will presumably keep the integral
(2.10) unchanged. If this argument should be valid,   our  path integral
formula in (2.4) would be  justified.  If our
speculation is correct, the formal path integral formula (2.5) will
provide a basis for the analysis of some non-perturbative aspects of QCD.

On the other hand, the Gribov problem may also suggest the presence of some
field configurations which do not satisfy any given gauge condition in four
dimensional non-Abelian gauge theory [2]. For example, one may not be able to
find any gauge parameter $\omega (x)$ which satisfies
\begin{equation}
\partial^{\mu}A_{\mu}^{\omega}(x) = 0
\end{equation}
for some fields $A_{\mu}$. Although the measure of such field configurations in
path integral is not known, the presence of such filed configurations would
certainly modify the asymptotic correspondence in
(2.11).  In the context of BRST symmetry, the Gribov problem may then induce
complicated phenomena such as the dynamical instability
of BRST symmetry[24]. If the dynamical instability of BRST symmetry should take
place, the relation corresponding to (3.27), which is a
result of the BRST invariance of the vacuum, would no longer be derived.  In
the framework of path integral, this failure of (3.27) would be recognized as
the failure of the expansion (3.19) since the normalization factor $N$ in
(3.21) would generally depend on not
only field variables but also $\lambda$ if the global single-valuedness in
(3.17) should be violated.

Finally, we note that the lattice gauge theory [25], which is based on
compactified field variables, is expected to change the scope and character of
the Gribov problem completely. The Gribov problem is
intricately related to the difficult issue of the non-perturbative continuum
limit of lattice gauge theory.[ Note that the perturbative continuum limit is
not quite relevant in the present context, since we do not worry about the
Gribov problem much in weak field perturbation theory].

\begin{figure}
\epsfbox{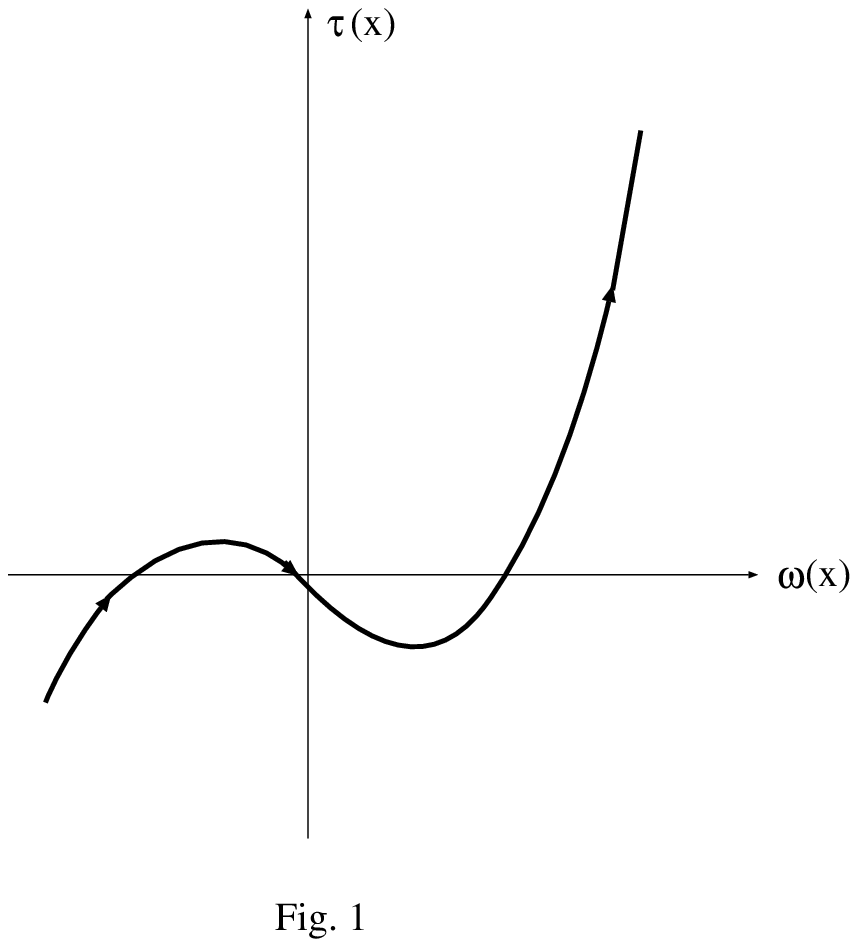}
\caption{ A schematic representation of eq. (2.11) for fixed
$A_{\mu}$.}
\end{figure}


\begin{thebibliography}{1}

\bibitem{1}
V. N. Gribov, Nucl. Phys. {\bf B139}(1978)1.
\bibitem{2}
I. M. Singer, Comm. Math. Phys. {\bf 60}(1978)7.
\bibitem{3}
R. Jackiw, I. Muzinich and C. Rebbi, Phys. Rev. {\bf D17}(1978)1576.
\bibitem{4}
T. Maskawa and H. Nakajima, Prog. Theor. Phys. {\bf 60}(1978)1526.
\bibitem{5}
K. Fujikawa, Prog. Theor. Phys. {\bf 61}(1979)627.\\
P. Hirschfeld, Nucl. Phys. {\bf B157}(1979)37.
\bibitem{6}
C. Becchi, A. Rouet and R. Stora, Comm. Math. Phys. {\bf 42}(1975)127;\ Ann.
Phys.{\bf 98}(1976)287.\\
M. Henneaux, Phys. Reports {\bf 126}(1985)1.\\
L. Baulieu, Phys. Reports {\bf 129}(1985)1.
\bibitem{7}
D. Zwanziger, Nucl. Phys. {\bf B412}(1994)657 and references therein.\\
L. J. Carson, Nucl. Phys. {\bf B266}(1986)357.\\
H. Yabuki, Ann. Phys. {\bf 209}(1991)231.\\
F. G. Scholtz and G. B. Tupper, Phys. Rev. {\bf D48}(1993)1792.
\bibitem{8}
R. Friedberg, T. D. Lee, Y. Pang and H. C. Ren, Columbia report,
CU-TP-689;\ RU-95-3-B.
\bibitem{9}
L. D. Faddeev and V. N. Popov, Phys. Lett. {\bf 25B}(1967)29.\\
G. 't Hooft, Nucl. Phys. {\bf B33}(1971)173.
\bibitem{10}
L. D. Faddeev, Theor. Math. Phys. {\bf 1}(1970)1.
\bibitem{11}
S. Coleman, The Uses of Instantons in {\em Aspects of Symmetry} (Cambridge
Univ. Press, Cambridge,1985).
\bibitem{12}
T. Kugo and I. Ojima, Phys. Lett. {\bf 73B}(1978)459.
\bibitem{13}
K. Fujikawa, Prog. Theor. Phys.{\bf 63}(1980)1364 ;\  ibid, {\bf 59}(1978)2045
\bibitem{14}
T. D. Lee and C. N. Yang, Phys. Rev. {\bf 128} (1962)885.
\bibitem{15}
K. Fujikawa, B. W. Lee and A. I. Sanda, Phys. Rev. {\bf D6}(1972)2923.
\bibitem{16}
E. S. Fradkin and I. V. Tyutin, Phys. Rev {\bf D2}(1970)2841.\\
A. A. Slavnov, Theor. Math. Phys. {\bf 10}(1972)99.\\
J. C. Taylor, Nucl. Phys. {\bf B33}(1971)436.
\bibitem{17}
P. A. M. Dirac, {\em Lectures on Quantum Field Theory}(Yeshiva Univ., New York,
1966).
\bibitem{18}
C. S. Lam, Nuovo Cim. {\bf 38}(1965) 1755.
\bibitem{19}
Y. Nambu, Prog. Theor. Phys. {\bf 7}(1952)131.
\bibitem{20}
J. D. Bjorken, Phys. Rev. {\bf 148}(1966)1467.\\
K. Johnson and F. E. Low, Prog. Theor. Phys. Suppl.{\bf 37-38}(1966)74.
\bibitem{21}
R. P. Feynman and A. R. Hibbs, {\em Quantum Mechanics and Path Integrals} (New
York, McGraw-Hill,1965).
\bibitem{22}
E. Witten, Nucl. Phys. {\bf B149}(1979)285.
\bibitem{23}
G. 't Hooft, Nucl. Phys. {\bf B72}(1974)461.
\bibitem{24}
K. Fujikawa, Nucl. Phys. {\bf B223}(1983)218.
\bibitem{25}
K. G. Wilson, Phys. Rev.{\bf D10}(1974)2445.


\end{thebibliography}
\end{document}